# Morphology of Buried Interfaces in Ion-Assisted Magnetron Sputter-Deposited $^{11}B_4C$-Containing Ni/Ti Multilayer Neutron Optics Investigated by Grazing-Incidence Small-Angle Scattering

Sjoerd Stendahl, Naureen Ghafoor, Matthias Schwartzkopf, Anton Zubayer, Jens Birch, and Fredrik Eriksson*

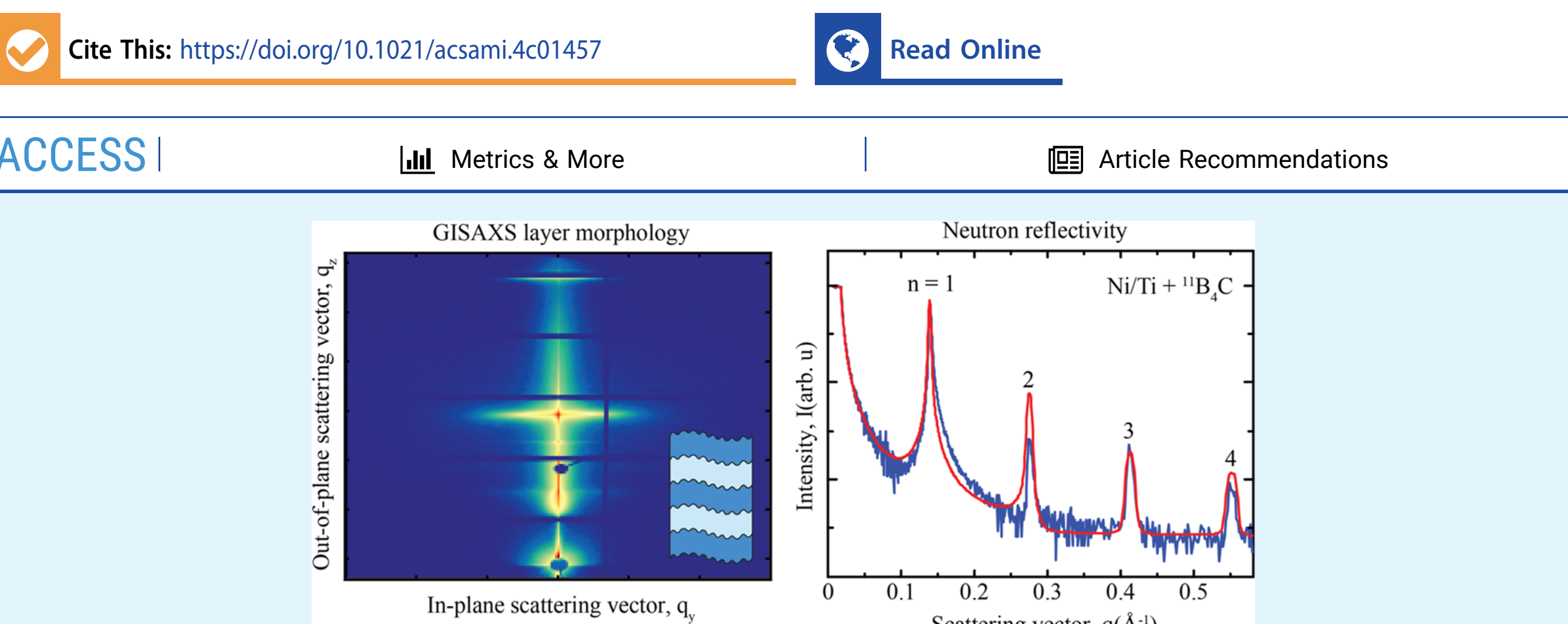

**ABSTRACT:** Multilayer neutron optics require precise control of interface morphology for optimal performance. In this work, we investigate the effects of different growth conditions on the interface morphology of Ni/Ti-based multilayers, with a focus on incorporating low-neutron-absorbing $^{11}B_4C$ and using different ion assistance schemes. Grazing-incidence small-angle X-ray scattering was used to probe the structural and morphological details of buried interfaces, revealing that the layers become more strongly correlated and the interfaces form mounds with increasing amounts of $^{11}B_4C$. Applying high flux ion assistance during growth can reduce mound formation but lead to interface mixing, while a high flux modulated ion assistance scheme with an initial buffer layer grown at low ion energy and the top layer at higher ion energy prevents intermixing. The optimal condition was found to be adding 26.0 atom % $^{11}B_4C$ combined with high flux modulated ion assistance. A multilayer with a period of 48.2 Å and 100 periods was grown under these conditions, and coupled fitting to neutron and X-ray reflectivity data revealed an average interface width of only 2.7 Å, a significant improvement over the current state-of-the-art commercial Ni/Ti multilayers. Overall, our study demonstrates that the addition of $^{11}B_4C$ and the use of high flux modulated ion assistance during growth can significantly improve the interface morphology of Ni/Ti multilayers, leading to improved neutron optics performance.

## ■ INTRODUCTION

Neutron scattering is a widely used experimental technique that enables nondestructive investigation of the structure and dynamics of various materials. This technique offers several advantages over conventional X-ray scattering, including sensitivity to light elements and the ability to probe the magnetic properties of materials. However, neutron scattering is limited by the relatively low flux of neutrons arriving at the experiment, which can reduce the overall signal-to-noise ratio and limit achievable resolution. Although the European Spallation Source (ESS) is expected to have the highest neutron peak flux in the world,[1] it will still fall short of the attainable flux for X-rays at synchrotron sources by several orders of magnitude. One significant factor contributing to the loss of neutron flux arriving at the experiment is the loss of neutrons in different optical components.[2]

Therefore, improving the performance of optical components such as multilayer neutron optics is a key strategy for enhancing the neutron flux at experiments. The conventional choice of materials for such multilayers is Ni/Ti, owing to the high contrast in scattering length density (SLD), which is a necessity for high reflectance.[3] However, the reflectivity performance of these optics is strongly dependent on the achievable interface width. To accurately account for the total loss of Bragg peak reflectivity in a periodic multilayer neutron mirror, it is necessary to introduce a Debye−Waller-like factor that describes the out-of-plane variation of the nuclear scattering potential by an error function

$$R = R_0 e^{-\left(2\pi m \frac{\sigma}{\Lambda}\right)^2} \quad (1)$$

where $R_0$ and $R$ indicate the reflectivity before and after taking the widths of the interfaces into account, $m$ describes the order of the Bragg peak, $\Lambda$ is the multilayer period, and $\sigma$ is the interface width. The interface width thus plays a critical role in determining the reflectivity performance of multilayer neutron mirrors. The interface width is influenced by two physically different factors, namely, the nonabruptness of the interface, due to interdiffusion and intermixing and the amount of interfacial roughness caused by factors such as nanocrystal facets and kinetically limited film growth. These two factors can not be distinguished in the specular direction and hence are considered as a single factor. The reflectivity performance of neutron mirrors is exponentially dependent on the interface width squared, as shown in eq 1. Therefore, even a small improvement in the interface width can have a significant impact on the reflectivity performance of the optical components.

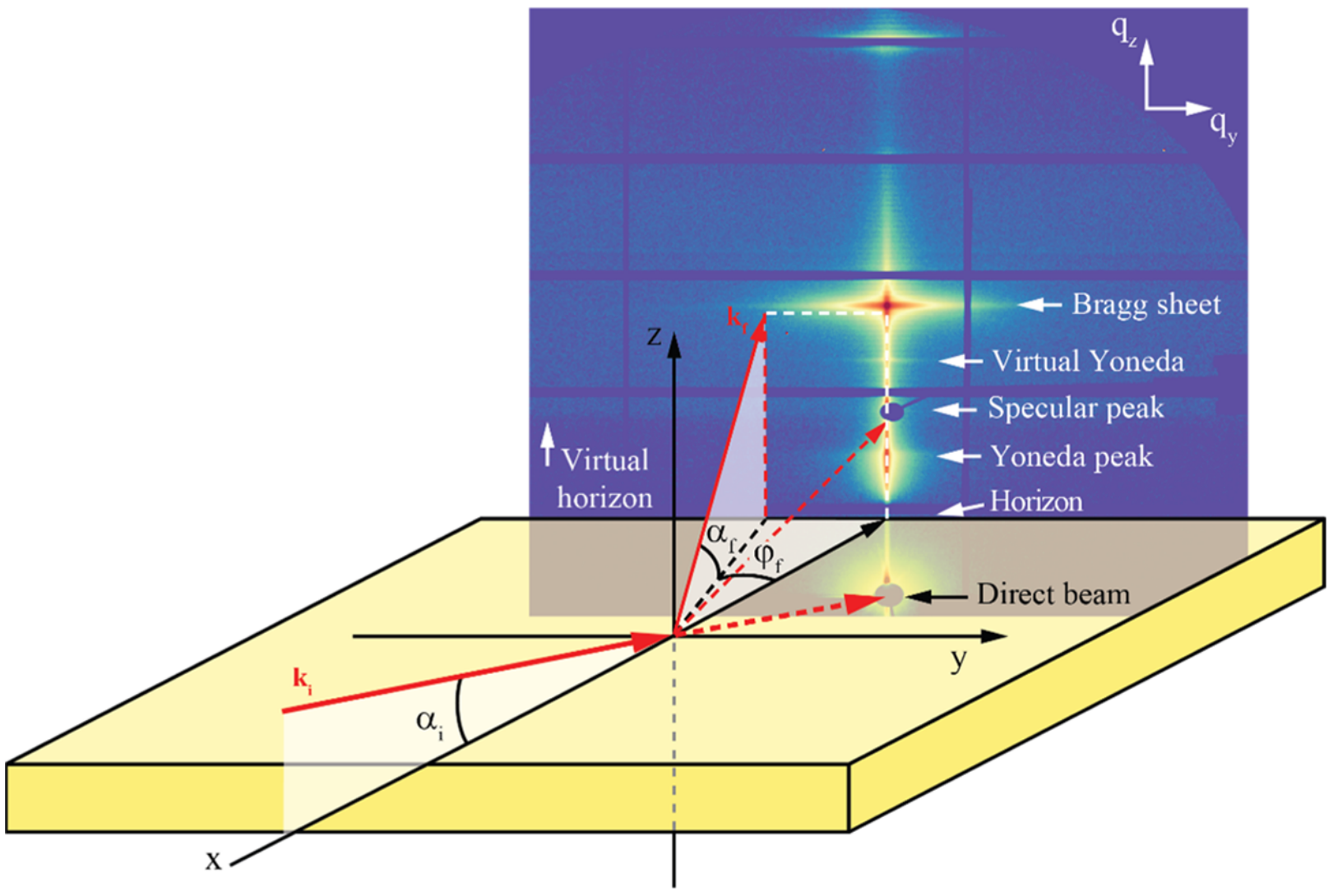

**Figure 1.** Schematic illustration of the GISAXS measurement geometry and a typical GISAXS reciprocal space map, highlighting distinct features. Information about the lateral morphology of the multilayer is found along the $q_y$-direction, while information about the vertical correlations in the multilayer is found along the $q_z$-direction.

Due to this reason, the majority of multilayer neutron optics research has been focused on improving the interface width. Currently, commercially available state-of-the-art neutron supermirrors are made from Ni and Ti and exhibit an interface width of 7 Å, as determined by simulated fits to specular neutron measurements.[4] There have been various approaches to improve the attainable interface width. For example, reactive magnetron sputter deposition of Ni in an Ar/air mixture resulted in improved interface width for thicker layers in a supermirror, as measured by neutron reflectivity.[5] Another approach involved adding an ultrathin Cr layer at the Ni/Ti interfaces to prevent interdiffusion between Ni and Ti. Neutron reflectometry confirmed a higher reflectivity as a result of lower interdiffusion.[6] Furthermore, an inherent asymmetry in surface free energy between Ni and Ti interfaces introduces both stress and extra roughness during multilayer growth. Depositing intermediate Ag layers at the interfaces reduces this asymmetry and shows smoother interfaces as confirmed by fits to the experimental neutron reflectivity data.[7] These efforts demonstrate the importance of improving the interface width in multilayer neutron mirrors to achieve a high reflectivity performance.

The largest contributors to the interface width of conventional Ni/Ti multilayers are a combination of nanocrystallite faceting, the formation of intermetallics at the interfaces, and intermixing and interdiffusion between the layers. To address these issues, co-sputtering of low-neutron-absorbing $^{11}B_4C$ during magnetron sputter deposition[8] has been employed to prevent crystallization and formation of intermetallics, promoting an amorphous multilayer structure.[9] However, this technique limits adatom surface mobility, resulting in rough layers and accumulating roughness evolution during growth. To promote surface diffusion while avoiding bulk diffusion during sputter deposition, ion-assisted growth has been used, which attracts ions from the process plasma to the growing film by using a substrate bias voltage. If the substrate bias voltage is carefully chosen, the ion assistance can provide sufficient energy to the surface atoms to migrate to energetically favorable positions, leading to smoother interfaces. It has been shown that an increased flux of ions with lower energy reduces layer mixing at the interfaces while the positive effects of enhanced adatom mobility remain.[10,11]

Using a magnetic coil at the substrate position, secondary electrons generated in the sputtering process can be guided to the substrate region, where they increase the plasma density by ionization. This increases the ion-to-adatom arrival ratio during growth, allowing atoms on the surface to be displaced, while keeping the ion energy low enough to minimize ion-induced forward knock-on intermixing.[10] However, the use of such continuous ion assistance can lead to intermixing between adjacent interfaces. Even if only surface atoms are displaced, the atoms being deposited will be mixed with the atoms of the underlying surface layer during the initial growth stage of each layer. To overcome this issue, a high flux modulated ion assistance scheme has been introduced, where the initial 3 Å of each layer was grown using a grounded substrate bias voltage to prevent ion bombardment, while the remainder of each layer was grown using a substrate bias voltage of −30 V to densify the initial layer and stimulate adatom mobility to achieve a smooth and flat top surface for the next layer.[10] Hence, the energy of the ions attracted from the plasma is modulated in a way that minimizes intermixing into the previous layer and then stimulates adatom mobility to create a smooth top surface before the deposition of the next layer.

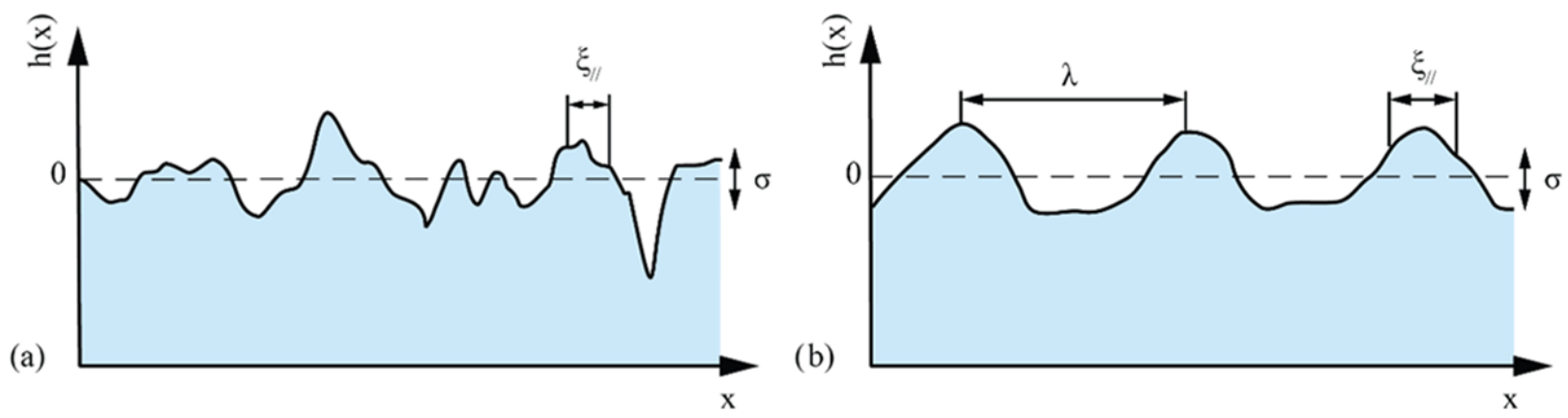

**Figure 2.** Illustration of (a) a self-affine interface with uncorrelated roughness and (b) a mounded interface with correlated roughness. The interface width is denoted by $\sigma$, the lateral correlation length is denoted by $\xi_{\|}$, and the mound separation is denoted by $\lambda$.

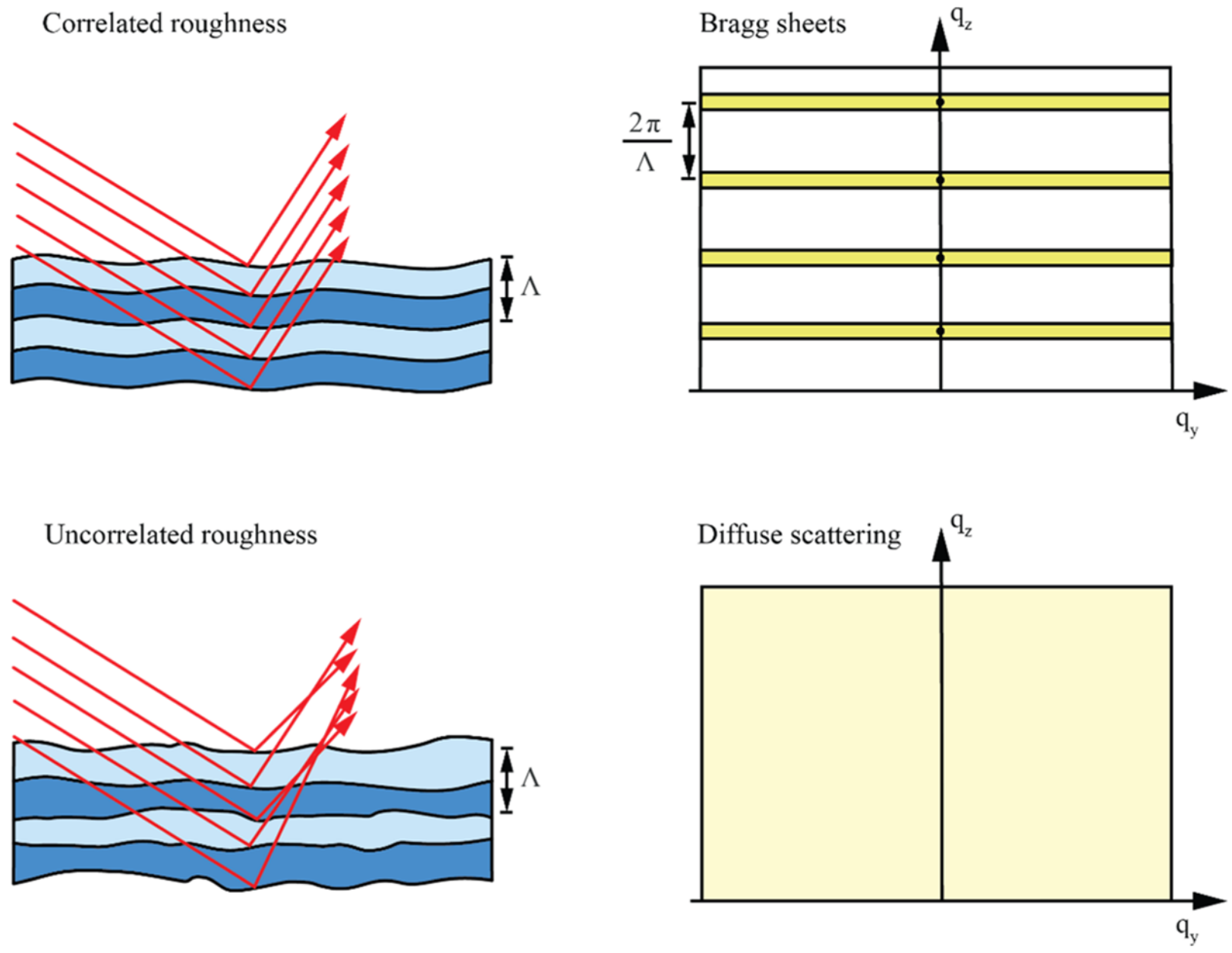

**Figure 3.** Illustration of correlated and uncorrelated interface roughness in multilayers and the resulting intensity distribution in reciprocal space. When the interface roughness is correlated, the diffusely scattered intensity is concentrated around the Bragg points, forming what are known as Bragg sheets. In contrast, uncorrelated interface roughness leads to a uniform intensity distribution throughout the reciprocal space.

To investigate the impact of adding $^{11}B_4C$ and using ion assistance during growth on the interface morphology of Ni/Ti multilayers, grazing-incidence small-angle X-ray scattering (GISAXS) was used in combination with neutron and X-ray reflectivity measurements and reflectivity fitting. First, the main features observed in GISAXS measurements from the multilayers are presented, along with the expected behavior for multilayer interfaces with lateral in-plane and vertical out-of-plane correlated or uncorrelated roughness. Next, the separate effects of $^{11}B_4C$ co-sputter deposition and ion assistance are discussed, and finally, the optimum conditions are compared, and the structural parameters using coupled reflectivity fitting are determined.

**GISAXS from Multilayers.** To achieve smooth and abrupt interfaces between the layers, it is crucial to have a good understanding of how different growth conditions affect the interface morphology in both the lateral and vertical directions. This has been studied using grazing-incidence small-angle X-ray scattering (GISAXS), and a typical experimental setup is shown schematically in Figure 1.

A typical GISAXS reciprocal space map is shown in Figure 1, where some distinct features are highlighted. At $q_z$-values corresponding to Bragg points ($q_z = (2\pi/\Lambda) \times n$, where $n$ is an integer), so-called Bragg sheets are formed due to correlated interface roughness.[12] The first-order Bragg sheet is highlighted in Figure 1. The Yoneda peak is caused by enhanced scattering at angles where the incident angle $\alpha_i$ is equal to the critical angle $\alpha_c$ of the multilayer. At this angle, the X-rays penetrate the top part of the multilayer, where the lateral components of the incident and reflected waves interfere constructively to form an evanescent wave, strongly enhancing the diffuse scattering. A beam stop is positioned below the horizon ($\alpha_f = -\alpha_i$) to block the incident direct beam transmitted through the sample. Another beam stop is positioned at $\alpha_f = \alpha_i$ to block the specularly reflected beam. At approximately the same $q_z$-values as the position of the specular peak, a sudden increase in background intensity can be observed, likely due to waveguiding effects resulting in a virtual horizon, which also gives rise to a second, virtual, Yoneda peak at an angle of $\alpha_{\text{virtual}} + \alpha_c$ from this position.

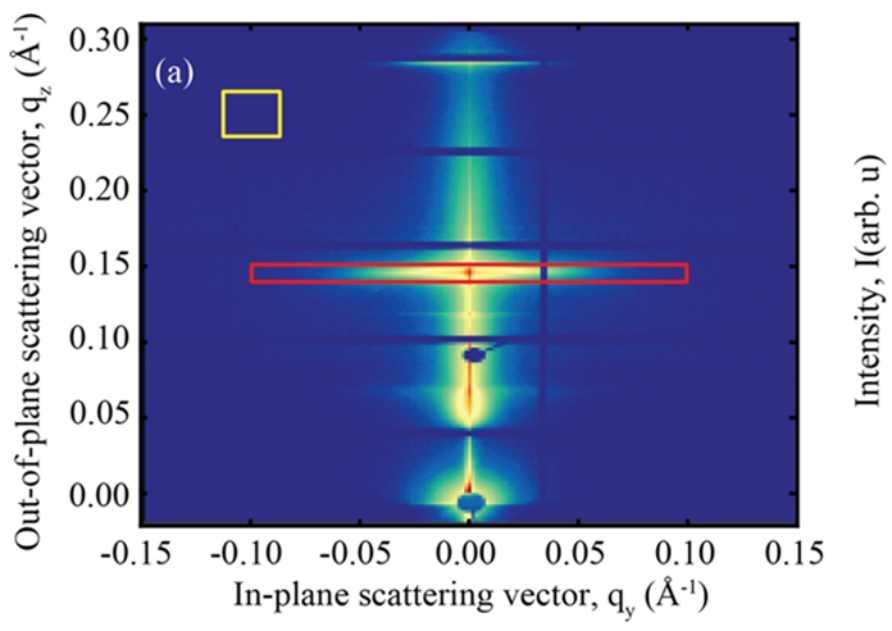

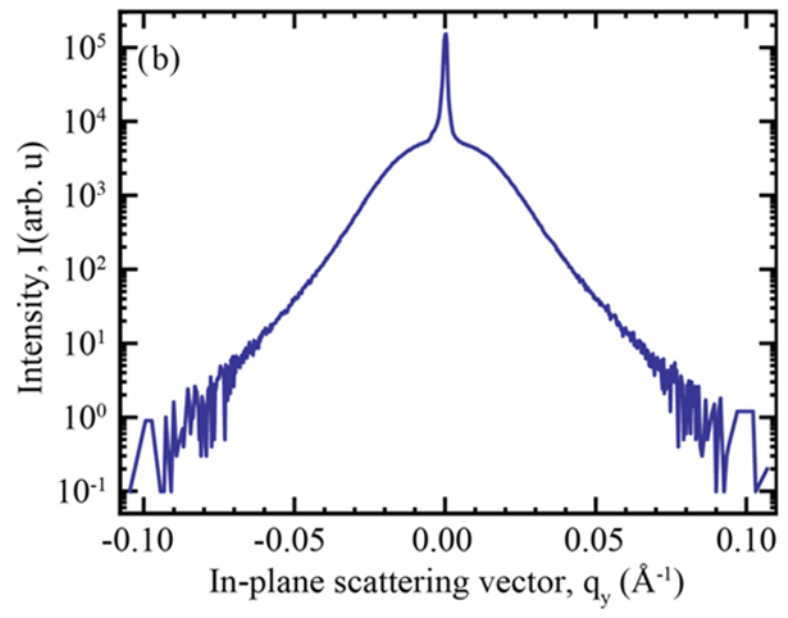

**Figure 4.** Results of a GISAXS measurement conducted in this study. (a) A 2D intensity map, with the red rectangle indicating the region used to generate an out-of-plane line scan at the first Bragg peak of the multilayer. The yellow rectangles show the region used to calculate the intensity for background subtraction. The vertical and horizontal lines in the scattering map with missing data correspond to physical wires passing through the 2D detector. These data points were removed from the retrieved intensity profiles and replaced with interpolated values. (b) Obtained out-of-plane line scan from the selected region of interest.

Two of the most important descriptions for interfaces in multilayers are that of self-affine and mounded interfaces. Self-affine interfaces exhibit a fractal-like nature where the exact shape of the interface will show the same features independent of the scaling level. Figure 2a shows a typical morphology of self-affine interfaces. For certain growth conditions, long-range periodic features occur at the interfaces in the form of mounds, as illustrated in Figure 2b. The distance between such mounds is often termed the wavelength and is commonly denoted by $\lambda$. To avoid confusion with the wavelength of the radiation source, the distance between these mounds is termed the mound separation in this work.

The autocorrelation function proposed by Sinha et al.[13] is the most commonly used method to describe self-affine interfaces. This function is expressed as an exponentially declining function given by

$$C(r) = \sigma^2 \exp\left[-\left(\frac{r}{\xi_{\parallel}}\right)^{2h}\right] \tag{2}$$

where $\xi_{\parallel}$ represents the lateral correlation length, signifying how quickly the autocorrelation function decreases. A high lateral correlation length indicates a relatively smooth surface, while h is the Hurst parameter, describing the jaggedness of the interface. The diffraction profile of the Bragg sheet along $q_y$ is the Fourier transform of this correlation function, revealing that a Hurst parameter of $h = 1.0$ results in a Gaussian interface profile, while a Hurst parameter of 0.5 corresponds to a Lorentzian interface profile. Any intermediate value must be solved analytically. Additionally, the lateral correlation length is inversely proportional to the obtained full width at half-maximum (FWHM) of the diffuse scattering curve.[14] In the case of mounded interfaces, this autocorrelation function requires modification using an oscillating term, commonly described by a first-order Bessel function[12,14]

$$C(r) = \sigma^2 \exp\left[-\left(\frac{r}{\xi_{\parallel}}\right)^{2h}\right] J_0\left(\frac{2\pi}{\lambda} r\right) \tag{3}$$

where $J_0$ is the first-order Bessel function and $\lambda$ is the distance between the mounds at the interfaces. In a multilayer system, the intensity scattered from multiple interfaces can add up constructively if their interface profiles are correlated. This phenomenon is illustrated in Figure 3, where correlated interfaces form distinct Bragg sheets of diffusely scattered intensity at specific $q_z$-positions, while completely uncorrelated interfaces scatter uniformly over $q_z$. To describe this cross-correlation function, Ming et al.[13] proposed the following function

$$C_{j,k}(r) = \sqrt{C_j(r) C_k(r)} \exp\left(\frac{-z_j - z_k}{\xi_{\perp}}\right) \tag{4}$$

Here, the $\sigma_j$ and $\sigma_k$ represent the interface widths of layer j and layer k, respectively, while $z_j - z_k$ describes the distance between the two layers. The vertical correlation length, denoted $\xi_{\perp}$, describes the length scale where the correlation function drops to a factor of $1/e$.

The vertical correlation length in a multilayer stack determines the degree of correlation between the layers and affects the width of the Bragg sheets in the $q_z$-direction. When the vertical correlation length is high, the layers are strongly correlated throughout the multilayer stack, resulting in narrow Bragg sheets. Conversely, when the vertical correlation length is significantly lower than the individual layer thickness, the interfaces are uncorrelated. From eq 4, it follows that the replication of the interface morphology is independent of the in-plane scattering vector $q_y$, which is in direct contradiction with experimental results, where the vertical correlation length is a function of the spatial frequency. A more realistic approach for the cross-correlation function follows from the Edwards–Wilkinson growth model, where low spatial frequencies of the interfaces are replicated better than higher ones.[12] Nevertheless, the rather simple Ming model is still applicable for a narrow range in $q_y$[14] and serves well as an illustration of how the vertical correlation length is related to the width of the scattered intensity.

One way to quantify the degree of correlation between interfaces is to use the effective number of correlated periods in the multilayer stack. This essentially describes the number of periods that fit within the vertical correlation length. This value is related to the FWHM of the Bragg sheet in the $q_z$-direction and can be expressed as

$$N_{eff} = \frac{2}{FWHM(q_y) \cdot \Lambda} \tag{5}$$

In this study, we have analyzed the shape of the first Bragg sheet for understanding the morphology of the multilayer interfaces. The FWHM of the Bragg sheet in the $q_z$-direction was determined at different $q_y$-positions to analyze the dependence of the vertical correlation length on the spatial frequency in $q_y$. The obtained values represent an average of the FWHM obtained at the positive and negative values of each $q_y$-coordinate to increase the statistics. To subtract the background scattering, the average intensity within a manually selected region was calculated and subtracted from that of each data point. The selected regions are illustrated in Figure 4a. Figure 4b shows the intensity variation across the first Bragg peak along the $q_y$-direction.

While the vertical scans provide information about the growth direction of the multilayer and the degree of correlation between interfaces, the scattering intensity along the $q_y$-direction reflects the lateral features of the interface morphology. The measured scattering intensity along the $q_y$-direction is directly related to the power spectral density (PSD) function of the interfaces, which corresponds to the Fourier transform of the autocorrelation functions. For a self-affine interface, the intensity profile decreases continuously with increasing lateral frequencies. In contrast, a mounded interface exhibits intensity in reciprocal space, corresponding to the mound separation in the interfaces. A shorter distance between the mounds, corresponding to a higher density of mounds, results in a peak at higher lateral frequencies and broader shoulders. The lack of a local maximum indicates a random distribution of interface mounds where no single characteristic length dominates.[15]

While GISAXS provides valuable information about the interface morphology, it does not directly reveal a clear value for the interface width. Intermixing, for example, reduces the intensity of the scattered signal instead of creating additional scattering effects, and a multilayer with stronger off-spectra scattering may have a lower interface width. Therefore, it is essential to combine GISAXS data with complementary techniques to obtain a comprehensive understanding of the interface profile. In this study, X-ray and neutron reflectivity measurements were used to compare the reflectivity performance between different multilayers, and the structural parameters of the multilayers were obtained from coupled fits to these data sets.

## ■ EXPERIMENTAL DETAILS

The multilayers in this study were deposited using a high vacuum magnetron sputter deposition system onto Si (001) substrates with a native oxide measuring 10 × 10 × 0.5 mm$^3$. Before growth, the background pressure of the chamber was reduced to $5.0 \times 10^{-7}$ Torr ($6.7 \times 10^{-6}$ Pa) using a turbo molecular pump backed by a rotary vane pump. The substrates were kept at ambient growth temperature and rotated at a constant rate of 17 rpm around the sample normal during growth to improve deposition uniformity and reduce shadowing effects. High-purity argon gas (99.9997%) was used as the sputtering gas at a pressure of 3 mTorr (0.4 Pa), as measured by a capacitance manometer.

The 3″ diameter sputtering targets for Ni and Ti were controlled in constant current mode using 80 and 160 mA currents, respectively. Co-sputtering of a 2″ diameter $^{11}B_4C$ target was performed by using a separate power supply in constant power mode. The power was varied between 0 and 70 W to achieve $^{11}$B+C concentrations in the range of 0−41.2 atom %, as determined by elastic recoil detection analysis (ERDA).

All multilayers in this study had a nominal period of Λ = 48 Å and a nominal layer thickness ratio of $\Gamma = d_{Ni}/(d_{Ni} + d_{Ti}) = 0.5$, with either $N$ = 50 or 100 periods. During growth, a negative bias voltage was applied to the substrate to attract ions from the plasma to the growing film surface. Two different ion assistance schemes were studied in this work: either a constant ion assistance scheme was used where a bias voltage of −30 V was applied to the substrate during the entire growth of the multilayer or a modulated ion assistance scheme with different voltages applied during different stages of the growth was used. For each layer deposited with the modulated bias scheme, an initial buffer layer of 3 Å was grown with a grounded substrate, while the rest of the layer was grown with a substrate bias voltage ranging from 0 to −50 V, depending on the sample. This modulated ion assistance design has previously proven to be successful in reducing roughness and eliminating intermixing in multilayer systems.[10,11] In addition to the applied substrate bias voltage, a magnetic coil surrounding the substrate position was used to guide secondary electrons generated in the sputtering process toward the substrate region, where they can ionize Ar sputter gas atoms. This increases the number of available Ar ions for ion-assisted growth and leads to an increasing ion-to-adatom arrival rate ratio, which displaces all adatoms on the surface while allowing for keeping the individual ion energy low enough to minimize ion-induced intermixing.[11]

The GISAXS measurements were carried out at the microfocus small- and wide-angle X-ray scattering beamline P03, located at the PETRA III synchrotron at DESY, Hamburg.[16] The X-ray beam was adjusted to a width of 25 μm and a height of 27 μm. A fixed incidence angle of $\alpha_i$ = 0.4°, just above the critical angle of 0.3° for the multilayer, was used along with a monochromatic wavelength of 0.96 Å. This combination provided a high-intensity X-ray beam with sufficient penetration depth to resolve the entire multilayer stack.[17] The scattered X-rays were collected using a PILATUS 2 M detector system, which comprised 1475 × 1679 pixels and had a field of view of 254 × 289 mm$^2$. The detector was positioned 3850 mm away from the sample, and a vacuum flight tube with a circular exit window of 200 mm diameter was placed in front of the detector. This resulted in a circular active area with a diameter of 1162 pixels. In reciprocal space, the scattering vector ranges covered were approximately $-0.15$ Å$^{-1}$ $< q_y <$ 0.15 Å$^{-1}$ and $-0.02$ Å$^{-1}$ $< q_z <$ 0.32 Å$^{-1}$. For each multilayer, a GISAXS measurement with a 1 s exposure time was performed at every $\Delta y$ = 0.5 mm step across the sample surface, perpendicular to the beam direction, which allowed us to confirm the sample uniformity. However, for the more detailed analysis, only the measurement at the center of each sample was used. The information was extracted using specialized open-source software, which was developed in-house for data reduction and analysis of these measurements.[18] All GISAXS line scans were obtained using line integration over a selected region of interest using this software.

X-ray reflectivity measurements were conducted using a Panalytical Empyrean diffractometer with a Cu X-ray tube. On the primary side, a parabolic X-ray mirror was used in combination with a 1/32° divergence slit, producing a parallel beam. On the secondary side, a parallel plate collimator was used in combination with a collimator slit, followed by a PIXcel-3D area detector in 0D mode. The reflectivity of the multilayers was measured in the range 0−10° 2$\theta$ with a step size of 0.01°/step and a collection time of 0.88 s per step, giving a total measurement time of approximately 30 min.

Neutron reflectivity measurements were conducted for a selection of multilayers at Institut Laue-Langevin in Grenoble, using the Swedish neutron reflectometer SuperADAM.[19] A monochromatic wavelength of 5.23 Å and a sample-to-detector distance of 150 cm were used for the measurements. The 2D detector at the reflectometer allowed for the simultaneous collection of both specular (along $q_z$) and off-specular (along $q_y$) reflectivity. Due to the higher acquisition intensity at lower incidence angles, the measurements were divided into four different regimes, with higher acquisition times at higher angles. In the first regime, a scan was performed from 0 to 4° 2$\theta$ with a step size of 0.02°/step and an acquisition time of 30 s per step. The second regime spanned from 2 to 4° 2$\theta$ using a step size of 0.02°/step and an acquisition time of 30 s per step. The third regime spanned from 8 to 16° $\theta$ using a step size of 0.04°/step and an acquisition time of 75 s per step, and the fourth and final regime spanned from 16 to 27° 2$\theta$ using a step size of 0.08°/step and an acquisition time of 160 s per step, adding up to a total acquisition time of approximately 15 h. To correct the footprint effect of the trapezoid-shaped beam, all measured multilayers were processed by using the dedicated data reduction software available at SuperADAM. The resulting intensities were then normalized at the critical angle.

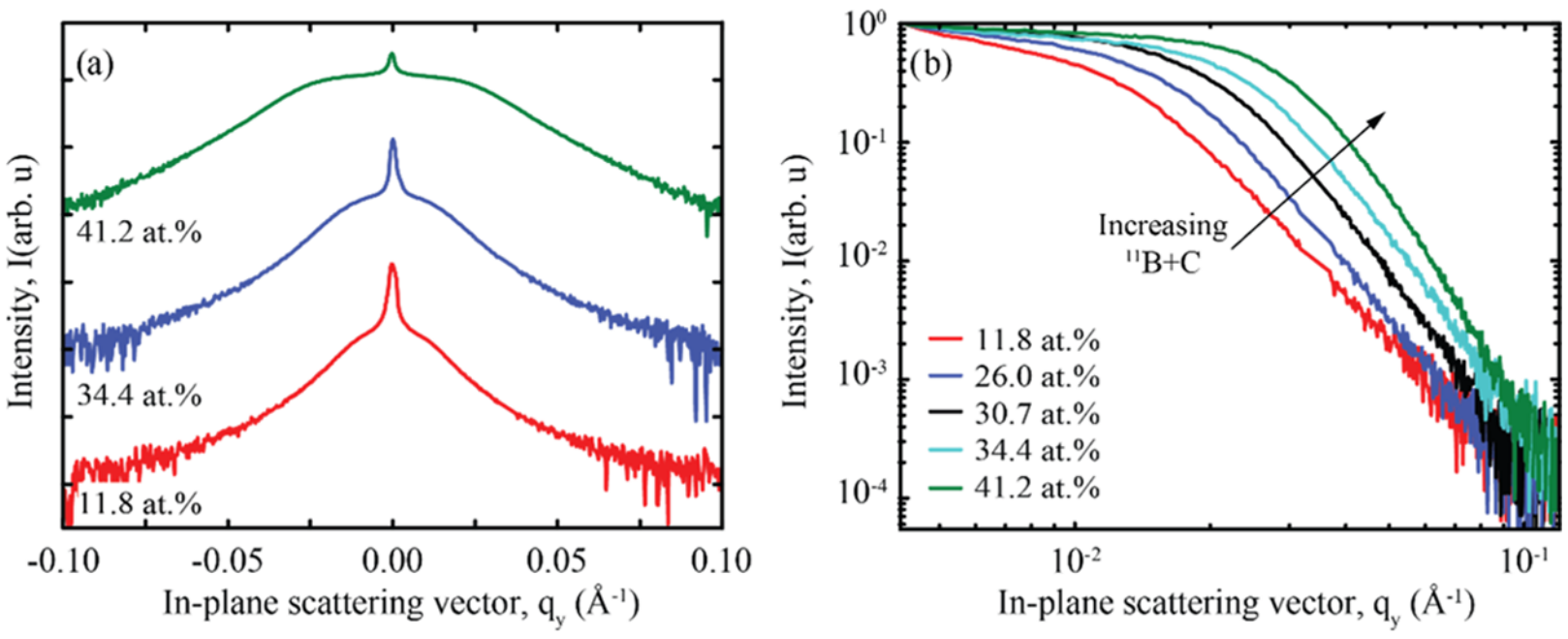

**Figure 5.** (a) Out-of-plane GISAXS line scans at the first Bragg sheet for different $^{11}$B + C concentrations. Note the log−log scale. (b) Complete out-of-plane line scans in the $q_z$-direction for a selected number of multilayers.

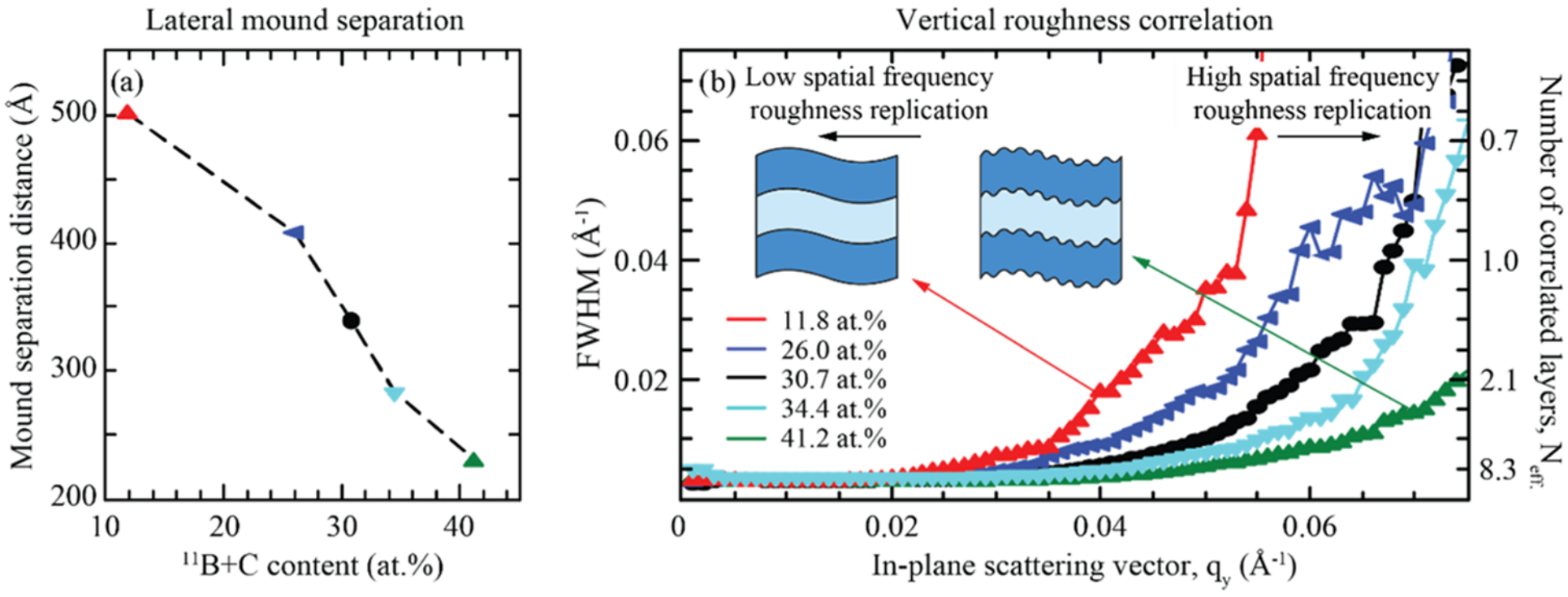

**Figure 6.** (a) Mound separation as a function of $^{11}$B + C concentration in the Ni/Ti multilayers. (b) FWHM in reciprocal space along $q_z$ at the first Bragg sheet at different positions in $q_y$, and the corresponding effective number of correlated layers for the same multilayers as in panel (a).

To determine the structure of the multilayers, including layer thicknesses, interface widths, and layer thickness drift, a single multilayer model, created within GenX reflectivity fitting software, was used.[20] X-ray and neutron reflectivity data were simultaneously fitted to this model, which includes a Parratt recursion formalism to calculate the specular reflectivity and an error function interface profile that accounts for interfacial roughness and intermixing. By coupling the structural parameters in the model, a single fit to the two independent data sets was obtained, which increased the fitting reliability.

The elemental composition of the films was analyzed by using time-of-flight elastic recoil detection analysis (ToF-ERDA) at the Tandem Laboratory at Uppsala University. A primary beam of $^{127}I^{8+}$ was used with an energy of 36 MeV at an incident angle of 67.5° relative to the surface normal. The energy detector was placed at a recoil scattering angle of 45°. The measured data were analyzed using Potku software to determine the atomic concentrations.[21] To account for the non-stochiometric ratio of $^{11}$B and C in the multilayer, the total sum of $^{11}$B + C was used instead of the concentration of $^{11}B_4C$. A detailed description of the experimental setup is given elsewhere.[22,23]

## RESULTS AND DISCUSSION

**Effect of $^{11}B_4C$ Co-deposition.** To investigate the impact of $^{11}B_4C$ co-deposition, a set of Ni/Ti multilayers were deposited with varying $^{11}B_4C$ magnetron powers, resulting in measured $^{11}$B + C concentrations of 11.8, 26.0, 30.7, 34.4, and 41.2 atom %, as measured using ToF-ERDA. The multilayers, having periods of 48 Å and a total of $N$ = 50 periods, were grown using the modulated ion assistance method.

Out-of-plane GISAXS line scans over the Bragg sheet show a clear formation of shoulders for all $^{11}$B + C concentrations, indicating the presence of mounded interfaces. As the concentration of $^{11}$B + C increased, the shoulders became broader and more pronounced, as shown in Figure 5a for multilayers 11.8, 34.4, and 41.2 atom %$^{11}$B + C concentrations. The characteristic length of the mounds at the interfaces can be estimated by finding the intersection between the tangents on either end of the shoulders in Figure 5b on a log−log scale. The resulting values, shown in Figure 6a, indicate that the characteristic length between the mounds decreased from approximately 500 Å at 11.8 atom %$^{11}$B + C concentration to less than 250 Å when the concentration is 41.2 atom %.

The reduction in adatom diffusion with an increasing amount of $^{11}$B + C in the Ni/Ti multilayers leads to a shorter separation between the mounds. The formation of mounded interfaces due to reduced surface diffusion is consistent with theories on nucleation and growth.[24,25] The shadow effects caused by island growth block the adatoms, resulting in undulation. The mounds acquire more adatoms and grow faster, while the valleys grow relatively slowly, exacerbating the shadowing effect. This effect further hinders the uniform growth of the following deposited layers and is known as roughening due to kinetically limited growth (often denoted by the misnomer "kinetic roughening").

Figure 6b illustrates the FWHM of the Bragg sheet along the $q_z$-direction for different lateral frequencies in the $q_y$-space. The graph reveals that an increase in the $^{11}$B + C concentration results in stronger correlations over a wider range of lateral frequencies. This is illustrated in the inset figures within Figure 6b, where the schematic multilayers are shown with vertical roughness correlations in only long spatial frequencies and with correlations in both long and short spatial frequencies. The multilayer shown on the right in Figure 6b schematically illustrates the case where a large amount of $^{11}$B + C is present in the sample, where, in addition to the replicated long-range features, also short-range features are replicated throughout the multilayer stack. This suggests that the incorporation of more $^{11}B_4C$ leads to interfaces that are more strongly correlated, with smaller spatial frequencies likely to repeat between subsequent layers. This behavior can be attributed to the reduced adatom diffusion resulting from the incorporation of $^{11}B_4C$. As the concentration of $^{11}$B+C increases, the total adatom diffusion decreases, promoting the formation of more interface mounds, which, in turn, shortens the distance between them.

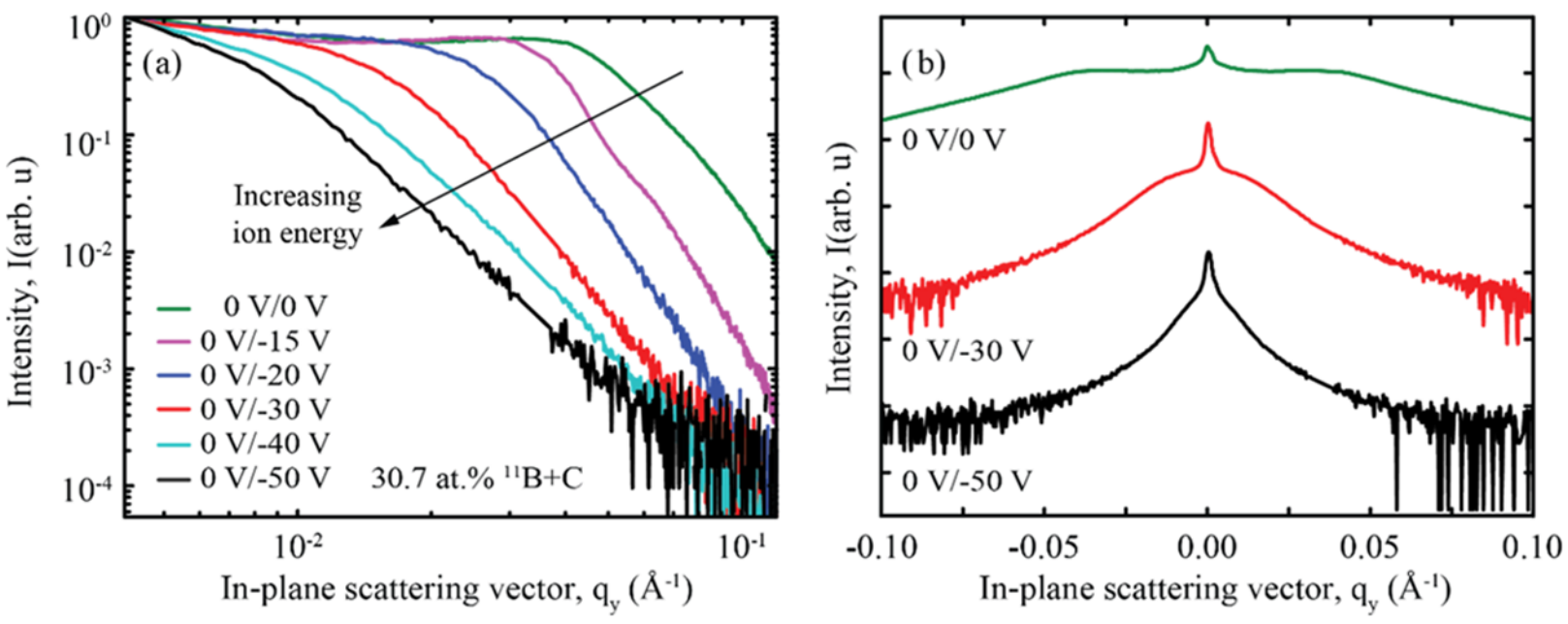

**Figure 7.** (a) Out-of-plane GISAXS line scans at the first Bragg sheet for $^{11}B_4C$-containing Ni/Ti multilayers grown using different substrate voltages during modulated ion-assisted growth, presented on a log−log scale. (b) Full out-of-plane line scans in the $q_z$-direction for a selected number of multilayers.

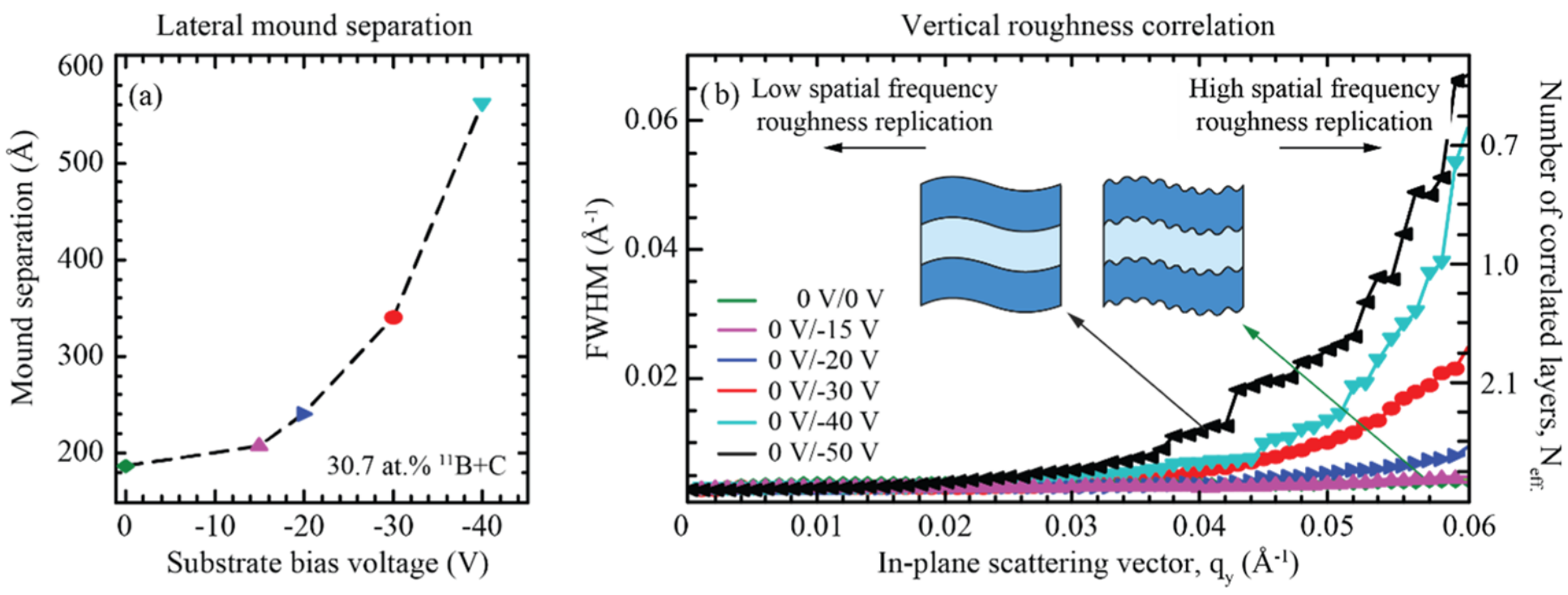

**Figure 8.** (a) Mound separation in $^{11}B_4C$-containing Ni/Ti multilayers as a function of the substrate bias voltage applied during modulated ion-assisted growth. (b) FWHM in reciprocal space in the $q_z$-direction at the first Bragg sheet for different positions in $q_y$ for multilayers grown with various substrate bias voltages.

**Effect of Ion Assistance.** To compensate for the reduction of adatom surface diffusion caused by the incorporation of $^{11}B_4C$, the effect of ion-assisted growth was explored. Specifically, the modulated ion assistance scheme involved growing the initial 3 Å of each layer using a grounded substrate for low ion energy bombardment and the final part of each layer using a higher substrate bias voltage for high ion energy bombardment. The substrate bias voltages were varied in the range 0 to −50 V while using a constant $^{11}B_4C$ magnetron power of 40 W, which corresponds to a concentration of 30.7 atom % of $^{11}$B + C for Ni/Ti multilayers with periods of 48 Å and a total of $N$ = 50 periods.

Figure 7a shows out-of-plane line scans on a log−log scale for $^{11}B_4C$-containing Ni/Ti multilayers grown by using different substrate voltages during modulated ion-assisted growth. The complete out-of-plane line scans in the $q_z$-direction for selected multilayers are shown in Figure 7b, for substrate 0, −30, and −50 V. When using voltages of 0 and 15 V, a distinct local maximum is visible at $q_y$ = 0.033 and 0.027 Å$^{-1}$, indicating that a specific lateral distance between the mounds dominates at these low ion energies. However, as ion energy increases, these local maxima vanish, and broad shoulders remain. At the highest ion energy, corresponding to a substrate bias voltage of −50 V, no shoulders remain, and the resulting curve closely resembles a self-affine interface.[26] Thus, these ion energies are sufficient to eliminate the mounds that form during the initial buffer layer growth at 0 V.

Figure 8a illustrates that mound separation at the interfaces increases with increasing ion energies, from approximately 200 Å separation without ion assistance to nearly 600 Å with a substrate bias voltage of −40 V. At a substrate bias voltage of −50 V, the intensity profile matches that of a self-affine interface, and there is no clear indication of interface mounds. Figure 8b shows that similar to the other measurements, the FWHM of the Bragg sheet broadens at higher spatial frequencies, indicating that low spatial features in real space are less likely to be replicated. At voltages below −20 V, the FWHM of the Bragg sheet is relatively constant over a wider range of lateral frequencies.

This aligns with expectations since these lower ion energies do not provide enough adatom mobility to produce surface diffusion for each added atom, resulting in short-range spatial features being replicated for subsequent layers. As the substrate bias voltage increases, the range in which the interface profile is vertically correlated between layers decreases. In other words, with higher ion energy, it becomes more difficult to replicate spatial features between subsequent layers.

The results of the GISAXS measurements show that increasing ion energy leads to a continuous improvement in the layer morphology with smoother interfaces. However, X-ray reflectivity data show that the quality of the multilayers no longer improves at substrate bias voltages higher than −30 V, suggesting that excessively high ion energies may activate bulk diffusion and, thus, cause mixing of the layers.

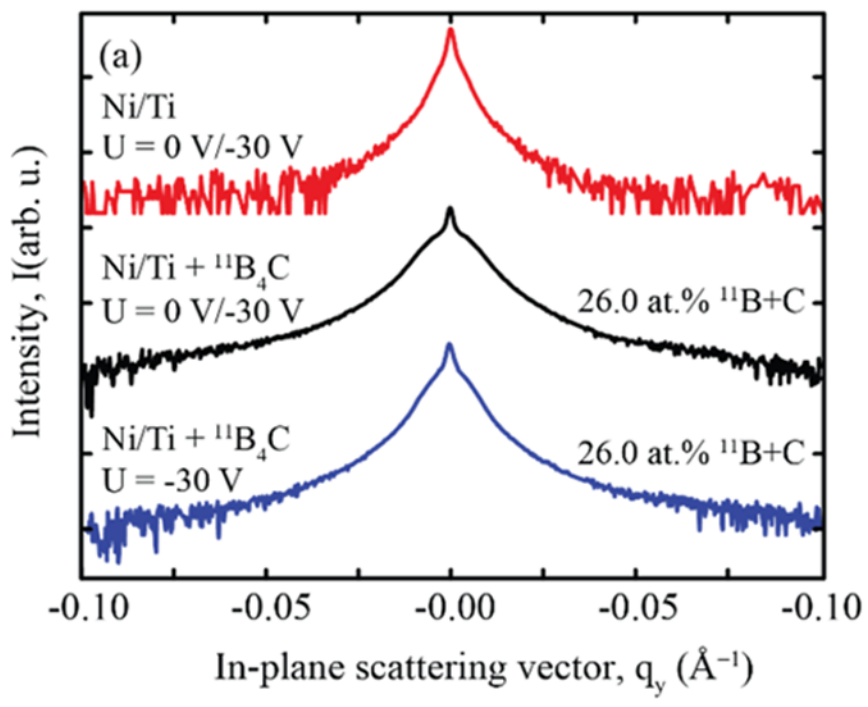

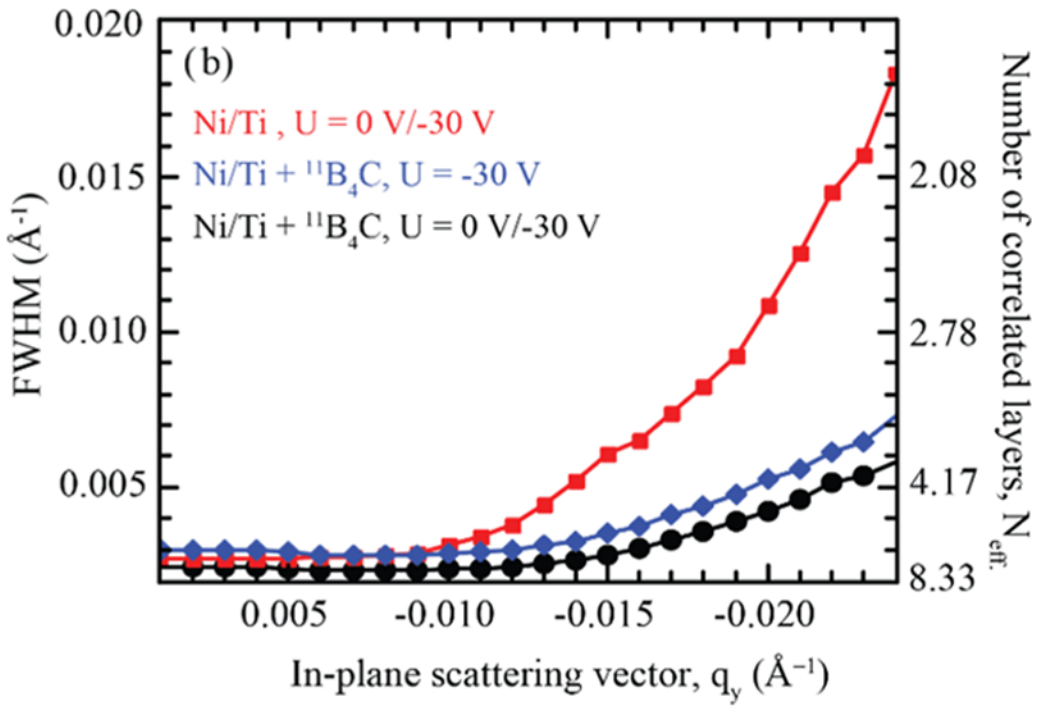

**Figure 9.** (a) Out-of-plane GISAXS line scans at the first Bragg sheet for three Ni/Ti multilayers. (b) FWHM of the first Bragg sheet in the $q_z$-direction is shown on the left-hand axis at different $q_y$-positions for the multilayers presented in panel (a). On the right-hand axis, the corresponding number of effectively contributing correlated bilayers, $N_{eff}$, is shown.

The incorporation of $^{11}B_4C$ and the use of ion assistance both have significant impacts on the interface morphology. The addition of $^{11}B_4C$ reduces the interface width in Ni/Ti multilayers by eliminating important contributors to a large interface width, but it also promotes the formation of interface mounds and leads to more strongly correlated layers. While a strong correlation between interfaces does not directly affect the specular reflectivity performance, it can cause roughness accumulation during growth. Furthermore, incorporating $^{11}B_4C$ in both layers reduces the optical contrast in terms of the scattering length density, which limits the potential reflectivity of the multilayer. Therefore, it is important to keep the amount of $^{11}B_4C$ to a minimum to achieve amorphous multilayers while reducing interface mounds and correlated interfaces and maintaining high optical contrast.

Ion assistance can counteract mound formation and correlated interfaces, but excessively high ion energies can eventually lead to intermixing. To prevent intermixing, an initial buffer layer is grown without ion assistance. However, GISAXS measurements show that the low adatom mobility during the formation of the buffer layer causes roughness that is not completely repaired when ion assistance is applied during the remainder of the layer growth.

**Comparison of the Effects of $^{11}B_4C$ and Modulated Ion Assistance.** This study shows that incorporating $^{11}B_4C$ is crucial for amorphizing the multilayers. However, it can also increase the correlation between interfaces and cause roughness accumulation due to reduced adatom mobility and surface diffusion.[10] To address this issue, a substrate bias voltage was applied during growth, as confirmed by GISAXS measurements. A modulated ion assistance scheme was used to overcome intermixing between subsequent layers, where the initial layer was grown with a grounded substrate bias, and the remaining layer was grown with an applied substrate bias voltage to allow for surface diffusion.

To compare the effects of these growth conditions on interface morphology, a series of multilayers consisting of $N$ = 100 periods, with a period of approximately 48 Å, were deposited. Figure 9a shows the out-of-plane line scans at the first Bragg peak for these multilayers. The top curve represents a pure Ni/Ti multilayer that was grown using a modulated ion assistance scheme at 0/−30 V. The middle curve shows a Ni/Ti multilayer with $^{11}B_4C$ co-deposited at a magnetron power of $P_{11B4C}$ = 30 W, corresponding to a total $^{11}B$ + C content of 26.0 atom %, and this multilayer was grown using a modulated ion assistance scheme at 0/−30 V. The bottom curve represents an $^{11}B_4C$-containing Ni/Ti multilayer with an $^{11}B$ + C content of 26.0 atom %, which was grown using constant ion assistance at −30 V.

Similar to the multilayers discussed earlier in this study, the resulting scans show clear broad shoulders around the center peak at $q_y$ = 0 when $^{11}B_4C$ is incorporated, indicating the presence of mounded interfaces throughout the multilayer. The multilayer grown with the modulated ion assistance scheme shows broader shoulders than the multilayer grown with a constant bias voltage, indicating a lower spatial distance between the mounds. As previously shown, a grounded substrate bias during multilayer growth results in a clear mounded interface with broad shoulders. However, the results presented in Figure 9a demonstrate that the rough and mounded interface profile of the initial buffer layer was not completely restored during the second growth stage, where a substrate bias voltage was applied.

The characteristic length, corresponding to the characteristic spacing between the mounds, has been obtained from these curves, resulting in a mound separation of 108 nm for the $^{11}B_4C$-containing multilayer grown with constant bias and 92 nm for the $^{11}B_4C$ containing multilayer with a modulated bias. These mounded interfaces can not be described using an exponentially declining autocorrelation function as described in eq 2, and simulations with a mounded interface model are beyond the scope of this paper. Nevertheless, the slightly lower characteristic length scale in the case with a constant bias voltage suggests that the multilayer grown with a modulated bias has a slightly higher density of mounds in the multilayer. This could indicate how the rough buffer layer that was grown during the first stage at grounded bias was not fully repaired during the second stage, where a substrate bias was applied.

Figure 9b shows the FWHM of the Bragg sheet along the $q_z$-direction at different lateral frequencies in the $q_y$-space for all three Ni/Ti multilayers. It can be observed that the Bragg sheet broadens at higher spatial frequencies, indicating that shorter vertical correlation lengths are not as clearly replicated between subsequent interfaces. Additionally, the multilayer grown with a constant bias voltage has a slightly broader Bragg sheet overall, compared to the modulated ion assistance scheme, suggesting that the latter gives rise to more correlated interfaces. This finding shows how the rough interfaces present in the initial buffer layer are not fully repaired during the second stage, where deposition occurs at a higher ion energy. For the pure Ni/Ti multilayer, it is mostly observed that the higher spatial frequencies are less correlated than those in the $^{11}B_4C$-containing multilayers. Thus, features at the interfaces with shorter spatial distances are less likely to be replicated in pure Ni/Ti multilayers than in $^{11}B_4C$-containing Ni/Ti multilayers.

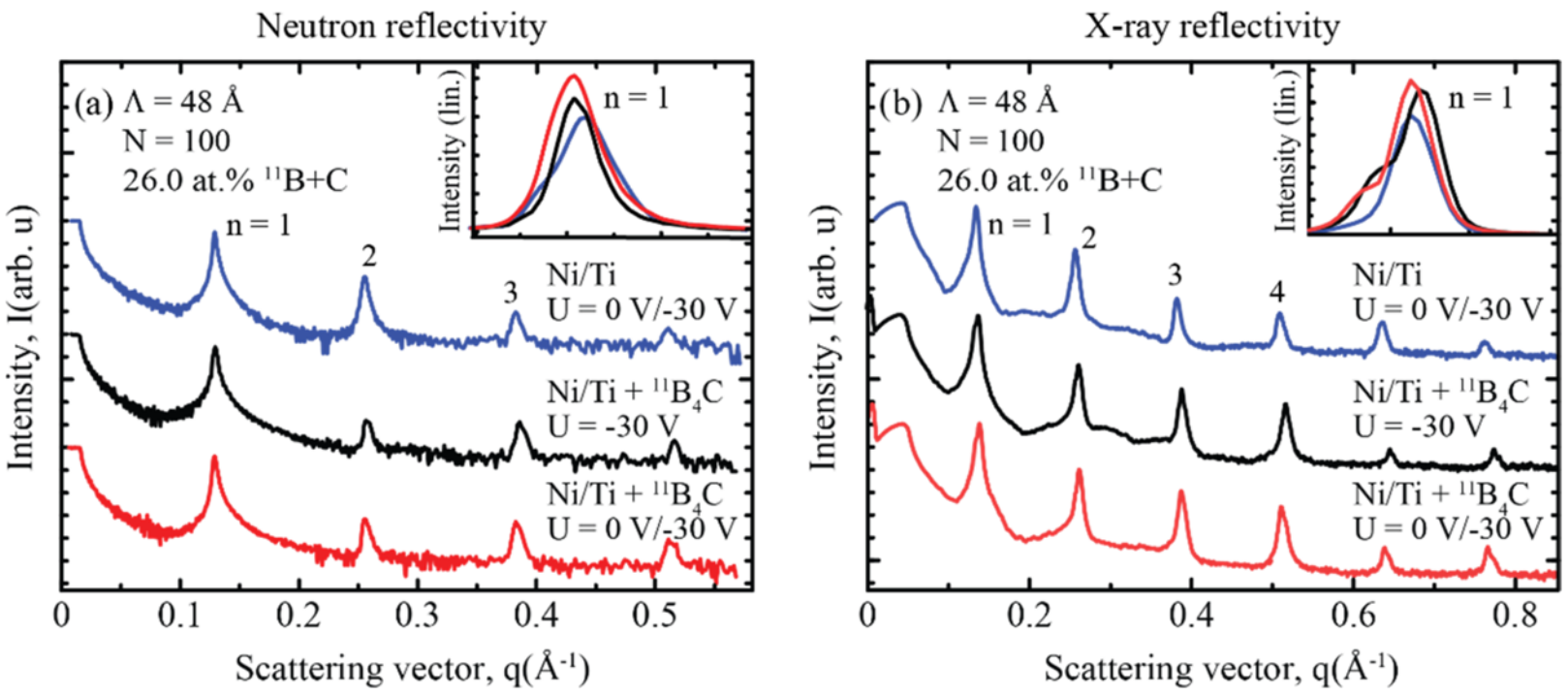

**Figure 10.** (a) Neutron reflectivity measurements for three multilayers grown using different ion-assistance conditions, with or without $^{11}B_4C$. (b) Corresponding X-ray reflectivity measurements. The reflectivity curves were shifted vertically for clarity. The inset graphs show the reflectivity at the first Bragg peak on a linear scale, where normalization was performed at the critical angle.

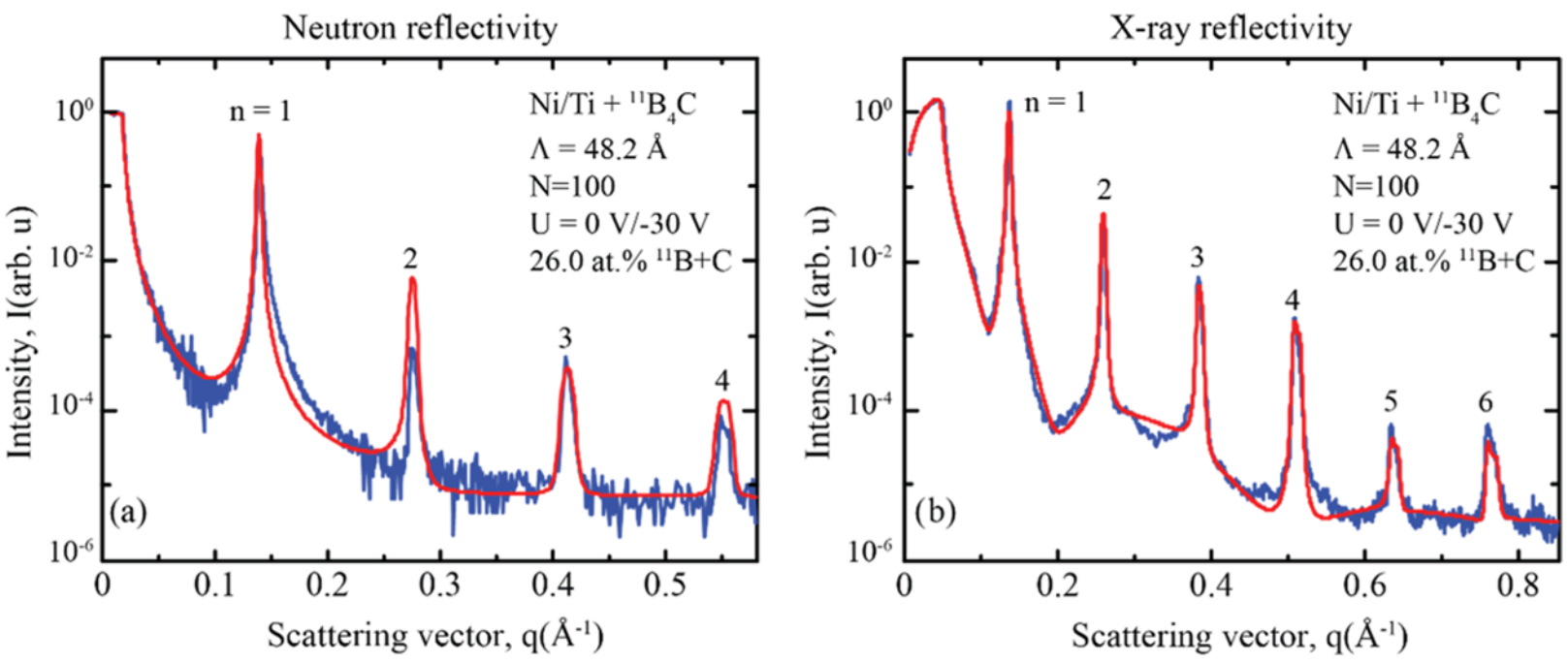

**Figure 11.** (a) Neutron and (b) X-ray reflectivity measurements and simulated fits. Both fits were performed simultaneously, with all structural parameters coupled to each other.

All three multilayers were measured with neutron and X-ray reflectivities, as shown in Figure 10. The results indicate an excellent reflectivity performance with clearly observable Kiessig fringes in the neutron data, indicating good interface quality. While the multilayer grown with both modulated ion assistance and $^{11}B_4C$ co-deposition shows the most correlated interfaces in the GISAXS data, both neutron- and X-ray reflectivity measurements reveal a clear increase in specular reflectivity for this multilayer compared to the other two cases. This illustrates a clear trade-off between roughness and intermixing. Although the pure Ni/Ti multilayer appears relatively smooth in the GISAXS signal with a low correlation between the interfaces, it is known that intermetallics tend to form between the interfaces, which is essentially a form of intermixing that will not show in the off-specular GISAXS data. Therefore, while the $^{11}B_4C$-containing multilayers do exhibit roughness in the form of interface mounds, the overall interface width is significantly improved, as demonstrated in previous work.[9,11] Comparing the $^{11}B_4C$-containing multilayers in these graphs, it is evident from Figure 9a that the multilayer grown with a modulated ion assistance scheme has a rougher interface with more interface mounds than the multilayer grown with a constant bias. Nevertheless, the modulated ion assistance scheme does lead to a significant improvement in terms of reflectivity, indicating a lower interface width.

Ultimately, the reduction of intermixing due to the initial buffer layer in the modulated ion assistance scheme outweighs the slight increase in roughness that is caused by this buffer layer.

The GISAXS data show that the inclusion of $^{11}B_4C$ results in interlayers with interface mounds that are strongly correlated. However, the neutron and X-ray reflectivity measurements demonstrate that the overall reflectivity performance is significantly improved by the incorporation of $^{11}B_4C$. These findings emphasize the significance of using an adequate amount of $^{11}B_4C$ to fully amorphize the layers without exceeding the required amount, which could lead to rough and strongly correlated interfaces. Furthermore, reducing the amount of added $^{11}B_4C$ is essential, as it decreases the scattering length density (SLD) contrast between the materials, thereby limiting the maximum achievable reflectivity.

**Reflectivity Performance.** X-ray reflectivity measurements revealed that the optimum reflectivity performance was achieved when adding 26.0 atom % of $^{11}B$ + C, which corresponds to an $^{11}B_4C$ magnetron power of 30 W, in combination with a modulated ion assistance using a substrate bias voltage of −30 V during the second stage. The resulting Ni/Ti multilayer was grown with a periodicity of 48.2 Å, consisting of 100 periods, and was measured using both neutron and X-ray reflectometry. To determine the structural parameters, the reflectivity curves were coupled and simultaneously fitted to a single model using GenX reflectivity software. The model allowed for linear increases in both the multilayer period and the interface width throughout the multilayer, taking possible thickness drifts and accumulated roughness into account. A detailed description of the fitting model can be found elsewhere.[9]

Figure 11 displays the results of the neutron and X-ray reflectivity measurements alongside their respective fits. The neutron reflectivity measurement shows four Bragg peaks, while the X-ray reflectivity measurement shows six Bragg peaks. The reflectivity fitting suggests that even more Bragg peaks would be visible if larger angular ranges had been measured. Due to the large number of periods in the multilayer and the limited angular resolution in both diffractometers, it is not possible to resolve the Kiessig fringes in these measurements.

Peak position fitting reveals that the multilayer period increases from 48.2 Å at the substrate to 48.8 Å at the top of the multilayer, corresponding to a total layer thickness drift of 0.0060 Å per period. Such a layer thickness drift may be caused by nonlinear deposition rates due to constantly eroding sputtering targets during growth, as well as a small drift in the timing accuracy of the computer-controlled shutters as the deposition progresses.

Table 1 summarizes the structural parameters obtained from the coupled neutron and X-ray reflectivity fitting. Although the resulting fits, in general, show excellent agreement with the experimental data with a figure of merit of only 0.136, there is a deviation at the second Bragg peak in the neutron reflectivity measurement. A better fit of this Bragg peak is not possible to achieve in this coupled fit without the introduction of additional fitting parameters. However, it is important to avoid adding additional fitting parameters, as it could lead to unphysical structural information. The deviation may be attributed to a slightly different effective interface width between the neutron and X-ray data. At such low values, only a minor difference in the interface width could significantly change the ratio between the interface widths, thus strongly influencing the reflectivity of the higher-order Bragg peaks.

**Table 1. Structural Parameters Obtained from Coupled Neutron and X-ray Reflectivity Fitting**[a]

| parameter | obtained value |
|---|---|
| initial multilayer period, Λ | 48.2 Å |
| thickness accumulation per period | 0.0060 Å |
| layer thickness ratio, Γ | 0.40 |
| initial interface width for the Ni + $^{11}B_4C$ layer | 3.0 Å |
| initial interface width for the Ti + $^{11}B_4C$ layer | 2.3 Å |
| accumulated roughness | 0.00 Å/period |
| figure of merit | 0.136 |

[a]The Figure of merit indicates how well the simulated model fits the neutron and X-ray experimental data on a log scale.

Based on the reflectivity fitting, it was observed that the $^{11}B_4C$-containing Ni layer has a slightly higher interface width (3.0 Å) compared to the corresponding one for Ti (2.3 Å). Similar asymmetric interface widths have previously been observed for pure Ni/Ti multilayers, which have been attributed to a strong Ni 111 texture leading to faceted crystallites at the top Ni interface, causing the roughness.[9,26] Despite the use of $^{11}B_4C$ to produce amorphous multilayers,[26] the asymmetry persists and could be due to the significant difference in surface free energy between Ni and Ti.[7] The reflectivity fits did not indicate any accumulation of the interface widths, consistent with ion-assisted depositions of $B_4C$-containing Ni/Ti multilayers in another deposition system.[8] This observation is particularly important for the production of supermirror optics, which may contain several thousands of layers.[27]

In particular, it can be noticed that the commonly reported average interface width is less than 2.7 Å, a significant improvement over the current state-of-the-art of 7.0 Å for commercially available Ni/Ti multilayer neutron optics.[4]

## ■ CONCLUSIONS

In this study, the effect of $^{11}B_4C$ co-deposition and ion assistance on the morphology of buried interfaces in Ni/Ti multilayer neutron optics have been investigated. Introducing $^{11}B_4C$ is known to amorphize the multilayers, which eliminates crystallite roughening and formation of intermetallics at the interfaces. Grazing-incidence small-angle X-ray scattering (GISAXS) analysis revealed that the layers become more strongly correlated and the interfaces form mounds with increasing amounts of $^{11}B_4C$. As the adatom mobility decreases, the characteristic separation between the mounds decreases, indicating an increase in the density of mounds for increasing amounts of $^{11}B + C$.

By applying high flux ion assistance during growth, the adatom mobility can be increased, reducing mound formation. However, this comes at the expense of a forward ion knock-on effect, which can lead to interface mixing. To prevent intermixing, a high flux modulated ion assistance scheme was used, where an initial buffer layer was grown with a low ion energy and the top of the layer with a higher ion energy. X-ray reflectivity measurements showed that intermixing is still possible if the applied ion energy is too high. Therefore, a careful balance between the different growth parameters is necessary to maximize the reflectivity potential.

The optimal condition was found to be adding 26.0 atom %$^{11}B + C$ combined with high flux modulated ion assistance. In each layer, the applied substrate bias voltage was initially kept grounded for 3 Å and then increased to −30 V. Such a multilayer, with a period of 48.2 Å and 100 periods, was grown, and the resulting structure was investigated by coupled fitting to neutron and X-ray reflectivity data. The average interface width was found to be only 2.7 Å, which is a significant improvement over the current state-of-the-art commercial Ni/Ti multilayers. These findings provide very promising prospects for high-reflectivity neutron optics, including periodic multilayers as well as broadband supermirrors for hot and epithermal neutrons.

## ■ ASSOCIATED CONTENT

### Data Availability Statement

Data underlying the results presented in this paper are not publicly available at this time but may be obtained from the authors upon reasonable request.

## ■ AUTHOR INFORMATION

### Corresponding Author

**Fredrik Eriksson** − *Department of Physics, Chemistry, and Biology, IFM, Linköping University, SE-581 83 Linköping, Sweden*; orcid.org/0000-0003-2982-5914; Email: fredrik.eriksson@liu.se

### Authors

**Sjoerd Stendahl** — *Department of Physics, Chemistry, and Biology, IFM, Linköping University, SE-581 83 Linköping, Sweden*

**Naureen Ghafoor** — *Department of Physics, Chemistry, and Biology, IFM, Linköping University, SE-581 83 Linköping, Sweden*

**Matthias Schwartzkopf** — *Deutsches Elektronen-Synchrotron DESY, Hamburg 22607, Germany;* orcid.org/0000-0002-2115-9286

**Anton Zubayer** — *Department of Physics, Chemistry, and Biology, IFM, Linköping University, SE-581 83 Linköping, Sweden*

**Jens Birch** — *Department of Physics, Chemistry, and Biology, IFM, Linköping University, SE-581 83 Linköping, Sweden;* orcid.org/0000-0002-8469-5983

Complete contact information is available at:
https://pubs.acs.org/10.1021/acsami.4c01457

### Notes

The authors declare no competing financial interest.

## ACKNOWLEDGMENTS

This research was funded by the Swedish Foundation for Strategic Research (SSF) within the Swedish national graduate school in neutron scattering (SwedNess) and the Swedish Research Council (VR). The Swedish Research Council VR Grant numbers 2019-04837, 2019-00191 (for accelerator-based ion-technological center in tandem accelerator laboratory in Uppsala University), and 2021-00357, Swedish Government Strategic Research Area in Materials Science on Advanced Functional Materials (AFM) at Linköping University (Faculty Grant SFO Mat LiU No. 2009 00971) are also acknowledged. The authors would like to acknowledge Peter Siffalovic at the Institute of Physics, Slovak Academy of Sciences, Slovakia, as well as Konstantin Nikolaev and Igor Makhotkin at XUV Optics at the University of Twente, The Netherlands, for fruitful discussions.